\documentclass[twocolumn]{svjour3}          
\smartqed  

\usepackage{graphicx}
\usepackage{graphicx}
\usepackage{algorithm}	
\usepackage{algorithmic}
\usepackage{verbatim}
\usepackage{amsmath}
\usepackage{amssymb}
\usepackage{setspace}
\journalname{Mobile Networks and Applications}

\begin{document}

\title{Renewable Energy Assisted Function Splitting in Cloud Radio Access Networks
\thanks{ This work is supported by the Turkish State Planning Organization (DPT) under the TAM Project, number 2007K120610.}
}


\author{Turgay Pamuklu \and
        Cicek Cavdar \and
        Cem Ersoy		
}


\institute{Turgay Pamuklu \at
              NETLAB, Department of Computer Engineering, Bogazici University, Bebek 34342, Istanbul, Turkey \\
              \email{turgay.pamuklu@boun.edu.tr}            \\
	    Cem Ersoy \at
              NETLAB, Department of Computer Engineering, Bogazici University, Bebek 34342, Istanbul, Turkey \\
              \email{ersoy@boun.edu.tr}           \\
	    Cicek Cavdar \at
             Communication Systems Department, KTH Royal Institute of Technology, Sweden \\
              \email{cavdar@kth.se}           \\
}

\date{Received: date / Accepted: date}
\maketitle

\begin{abstract}
Cloud-Radio Access Network (C-RAN) is a promising network architecture to reduce energy consumption and the increasing number of base station deployment costs in mobile networks. However, the necessity of enormous fronthaul bandwidth between a remote radio head and a baseband unit (BBU) calls for novel solutions. One of the solutions introduce edge-cloud layer in addition to the centralized cloud (CC) to keep resources closer to the radio units (RUs) and split the BBU functions between the center cloud (CC) and edge clouds (ECs) to reduce the fronthaul bandwidth requirement and to relax the stringent end-to-end delay requirements. This paper expands this architecture by combining it with renewable energy sources in CC and ECs. We explain this novel system and formulate a mixed integer linear programming (MILP) problem which aims to reduce the operational expenditure of this system. Due to the NP-Hard property of this problem, we solve it by using a MILP Solver and provide the results in this paper. Moreover, we propose a faster online heuristic to find solutions for high user densities. The results show that make splitting decisions by considering renewable energy provide more cost-effective solutions to the MNOs.
\end{abstract}

\section{Introduction}
\par In a distributed RAN architecture, a base station (eNodeB) is a combination of an analog radio unit and a digital baseband processing hardware (BBU). In a cloud RAN (CRAN) architecture, the functions of BBUs are operated in a centralized cloud, and they serve to more than one analog radio units which are called remote radio heads (RRHs) in this architecture. These two apart locations are connected to each other with high-speed links called fronthaul links \cite{Pfeiffer2015}. This new architecture has several benefits, such as energy-efficiency and ease of maintenance. On the other hand, it has disadvantages such as increased end-to-end delay and high-bandwidth requirement in optical fronthaul links, which are called common public radio interface (CPRI), between RRHs and BBUs \cite{Dotsch2013}. Therefore, the functional splitting of BBUs is proposed for 5G networks to eliminate these disadvantages \cite{3GPP2017,SmallCellForum2016}.
\par Although functional splitting of BBUs is a relatively new idea, there are several studies focusing on this problem. Mharsi et al. model the problem in a graph in which BBU functions are the nodes of the graph and the connections between these functions are the edges of this graph. Then they propose a greedy heuristic to minimize a multi-objective function which is the combination of the CPU core consumption and the end-to-end latency \cite{Mharsi2018}. Liu et al. also model the problem as a graph, but they prefer to use a genetic algorithm to find a solution \cite{Liu2015}. Checko et al. calculate the multiplexing gains of centralized BBUs, which comes from servicing the multiple cells that have low traffic loads, for different splitting decisions by using a teletraffic approach and simulation analysis \cite{Checko2016}. Wang et al. introduce a new approach to the function splitting concept, in which some of the functions are processed in an edge cloud (EC) as an alternative to the central cloud (CC). That edge may additionally serve more than one RRHs \cite{Wang2017}. They call this architecture "Hybrid CRAN", and they formulate the problem as a  multi-objective MILP problem, in which energy consumption and midhaul bandwidth are the two parts of this minimization problem. Also, in an additional study, they focus on end-to-end latency of this type of architecture, and they add the latency as a constraint in their problem \cite{Alabbasi2017}.
\par Several researchers already studied using renewable energy sources in the distributed RAN architecture. Some studies focus on the problem of distributing renewable energy sources among the base stations in an efficient way to increase the usage of renewable energy in a RAN \cite{Chia2014,Reyhanian2015,Ahmed2018}. The other aims of these papers include reducing the brown energy consumption and carbon emission rates, and traffic load and energy balancing between the base stations \cite{Ahmed2018a}. Moreover, some studies concentrate on sizing the renewable energy systems and operating the base stations effectively by using these systems \cite{son2011energy,Pamuklu2013,hanoptimizing,Pamuklu2018}. The problems they attempt to solve are NP-Hard problems, so they introduce several heuristics and methods to reduce the total cost of ownership (TCO) of a mobile network operator (MNO).
 \par Using renewable energy sources in a CRAN architecture is a recent research area.  Alameer et al. model this architecture as a queuing system. They deploy renewable energy sources in each RRHs and BBUs and focus on minimizing the overall energy consumption by considering the QoS \cite{Alameer2016}. Guo et al.  focus on a similar problem, in which they represent the system as an MINLP problem. They propose a two-phase heuristic to reduce the brown energy consumption \cite{Guo2018}. 
\par We propose a Green Hybrid CRAN model, in which renewable energy systems are deployed on each ECs and CC. We also provide an optimization problem which aims to reduce the brown energy consumption in this CRAN network by proper functional splitting decisions. The main contributions of this paper are as follows:
\begin{enumerate}
\item While there are separate studies that focus on functional splitting and using renewable energy sources in the CRAN, this paper is the first study that combines these two problems in the CRAN architecture.
\item We formulate an online problem which aims to reduce the operational expenditure (OpEx) of this new architecture. We present that in this network, we have to make decisions to split functions by considering to use renewable energy sources efficiently.
\item While MILP solvers produce reasonable solutions for this problem in a small RAN, they are not suitable for large RANs. Thus, we introduce a heuristic approach for larger RANs.
\item We test our methods for different traffic loads and solar radiation rates. Besides, we use real solar data and test our methods for different seasons to see the effect of seasonal changes in four different geographical areas in the world. We demonstrate the performance of our method in different locations with significantly different solar radiation patterns.
\end{enumerate}
\par The remainder of this paper is organized as follows. We describe this new type of CRAN and its cost optimization problem in the second and third sections, respectively. In the fourth section, we explain a heuristic approach for large RANs. In the fifth section, we provide the results, and in the last chapter, we conclude the paper.

\section{Green Hybrid Cloud Radio Access Network Model}
\par We can briefly explain the model in three main sections: the network architecture, the traffic model and the energy model.
\begin{figure} 
\includegraphics[width=0.5\textwidth]{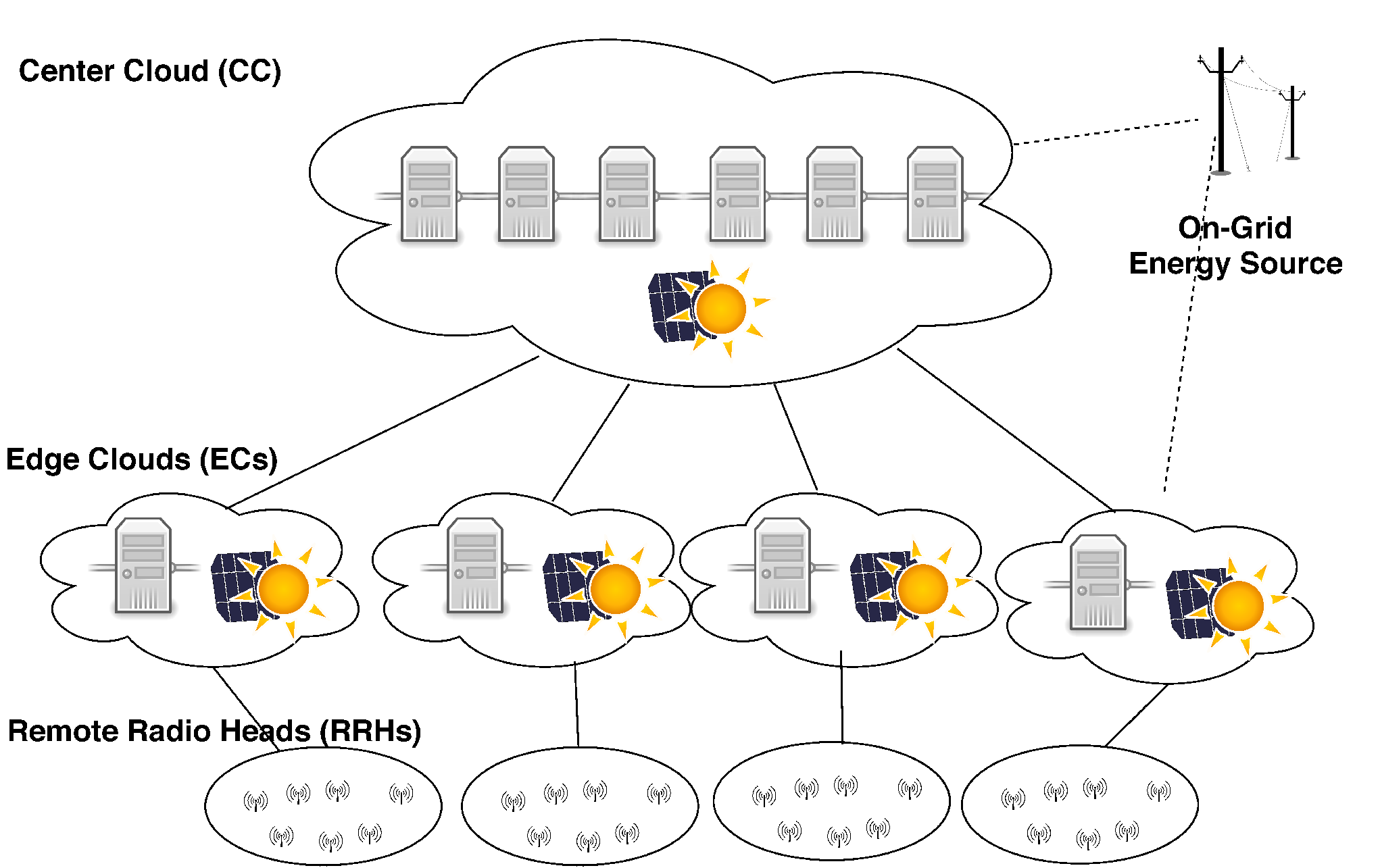}
\caption{\label{fig:renhcran} Green Hybrid Cloud Radio Access Network.}
\end{figure}  

\subsection{Network Architecture}
\par Figure~\ref{fig:renhcran} shows the network architecture.  In a classical RAN, BBU functions physically exist in one place. In this architecture, we break the chain of these functions in two certain split points and perform them in three layers, which are CC, ECs and RRHs. We may divide these functions into two main groups, cell related functions (CRFs) and user-related functions (URFs) \cite{Wang2017}. In this architecture, the chain of cell-related BBU functions may break between the RRH and EC, and they have operated either in an RRH or in an EC which depends on the decisions of the MNO. According to Small Cell Forum, splitting after CRFs reduces the required midhaul bandwidth significantly \cite{SmallCellForum2016,Maeder2014}. Therefore, to reduce the required midhaul bandwidth, the CC does not process CRFs in our network architecture. Besides, splitting after CRFs have multiplexing benefits because the interface and processing rates for URFs depend on the user traffic loads\cite{Dotsch2013,Checko2016}.   
\par
\begin{figure}
\centering
\includegraphics[width=0.48\textwidth]{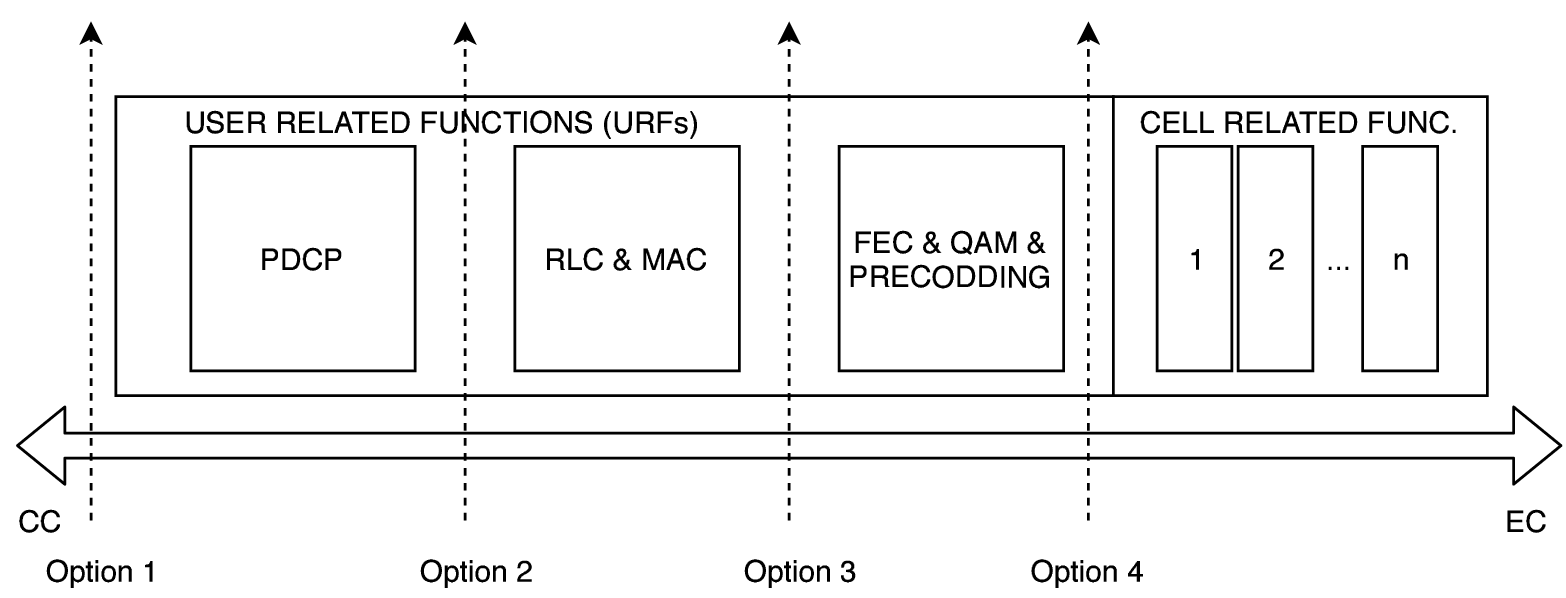}
\caption{\label{fig:splitting} Splitting Options.}
\end{figure}

 URFs may operate in the ECs or the CC, and this paper concentrates on these decisions. We have four different splitting options in the CC (Figure~\ref{fig:splitting}) :
\begin{enumerate}
\item CC does not process any URF.
\item CC processes only PDCP (Packet Data Convergence Protocol).
\item CC processes RLC (Radio Link Control) and MAC (Medium Access Control) in addition to the PDCP which means it processes all URFs higher than the physical layer.
\item CC processes FEC, QAM and precoding processes which means that it processes all URFs.
\end{enumerate}
\par Despite of these chosen splitting options, MNOs may modify them according to their networks. Roughly, if we process more URFs in the CC, we can reduce the energy consumptions, but we increase the need of midhaul bandwidth and end-to-end latency \cite{Checko2016}.

\subsection{Traffic Model}
\label{sec:userTraffic}
\begin{figure}
\centering
\includegraphics[width=0.48\textwidth]{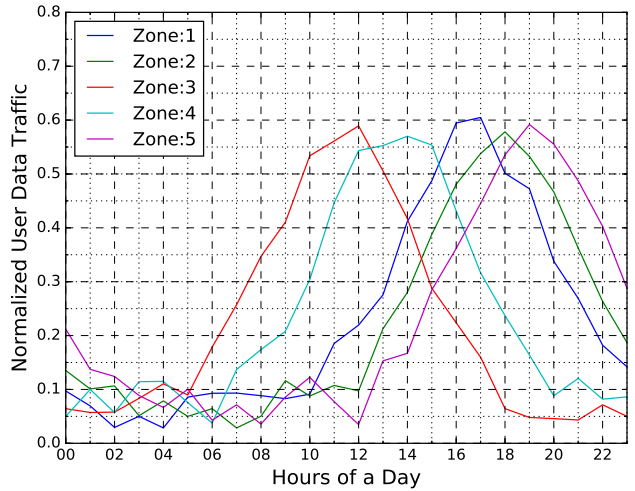}
\caption{\label{fig:traffic_day.pdf} Five different data traffic patterns in a day period.}
\end{figure}   

\par Peng et al. discover that the data traffic significantly varies both temporally and spatially in urban zones \cite{Peng2014}. Besides, they grant two critical conclusions that apply to our model. The first one is that the temporal variation of data traffic is more powerful in a day period, but it is not significant between the following days. Therefore, the historical traffic data can be used as input data for online splitting decisions. The second is that the data traffic loads are distinct among nearby locations, especially in their peak hours. For that reason, we generate different data traffic patterns for the users of different ECs.  
\par The temporal traffic profile we apply in this paper evolves from the previous related work. First, Marsan et al. \cite{Marsan2010} stated a formula which creates one-day sinusoidal shape traffic for a RAN. Next, Hossain et al. \cite{Hossain2010} revised this formula by adding a stochastic fluctuation in a day period. Finally, Zhang et al. \cite{Zhang2013} added a multiplier into this formula to create diversity between different locations. We enhance this last formula by adding a fluctuation between the days of four seasons. Our traffic pattern is given in Equation~\ref{eq:trafficCreator} in which $\varphi$ is a random value between the $3\pi/4$ and $7\pi/4$ which defines the peak hour of the traffic profile,  $\nu$ determines the slope of the traffic profile and $n(t)$ is a random value which produces a fluctuation in this traffic profile. Therefore, we can create the variation of the data traffic between each hour by this formula. 
\begin{equation}
\label{eq:trafficCreator}
f_{r}(t) = \frac{1}{2^{\nu}}[1+\sin(\pi t/12 + \varphi )]^{\nu} + n(t)
\end{equation}
\begin{equation}
\label{eq:meanArrivalRate}
\lambda_{it} = f_{r}(t),    i\in\mathcal{I}_{c},  c\in\mathcal{C}_{r}\quad [in \: bps]
\end{equation}
\begin{equation}
\label{eq:userTraffic}
\rho_{it} = \lambda_{it} / \mu_{it} \quad [in \: bps]
\end{equation}
\par We use a three-step approach to create a traffic diversity between the ECs according to the findings in \cite{Peng2014}. In the first step, we create five different traffic profiles using Equation~\ref{eq:trafficCreator} in which each profile have different peak hours (Figure~\ref{fig:traffic_day.pdf}) and allocate them to each EC. In the second step, the users of an EC demand the data traffic according to their EC traffic profile. We provide this assignment in Equation~\ref{eq:userTraffic}. Therefore we create a spatial diversity between the five different ECs which have different peak hours (Equation~\ref{eq:meanArrivalRate}). We use this calculated value $\lambda_{it}$ as a mean arrival rate of the users serviced by an EC. Finally, we create the traffic loads of the users by Equation~\ref{eq:userTraffic}. In this equation, user inter-arrival times have an exponential distribution, and the requested data size is an exponentially distributed random variable with mean $1/\mu_{it}$ \cite{oh2013dynamic}.

\begin{figure}
\centering
\includegraphics[width=0.48\textwidth]{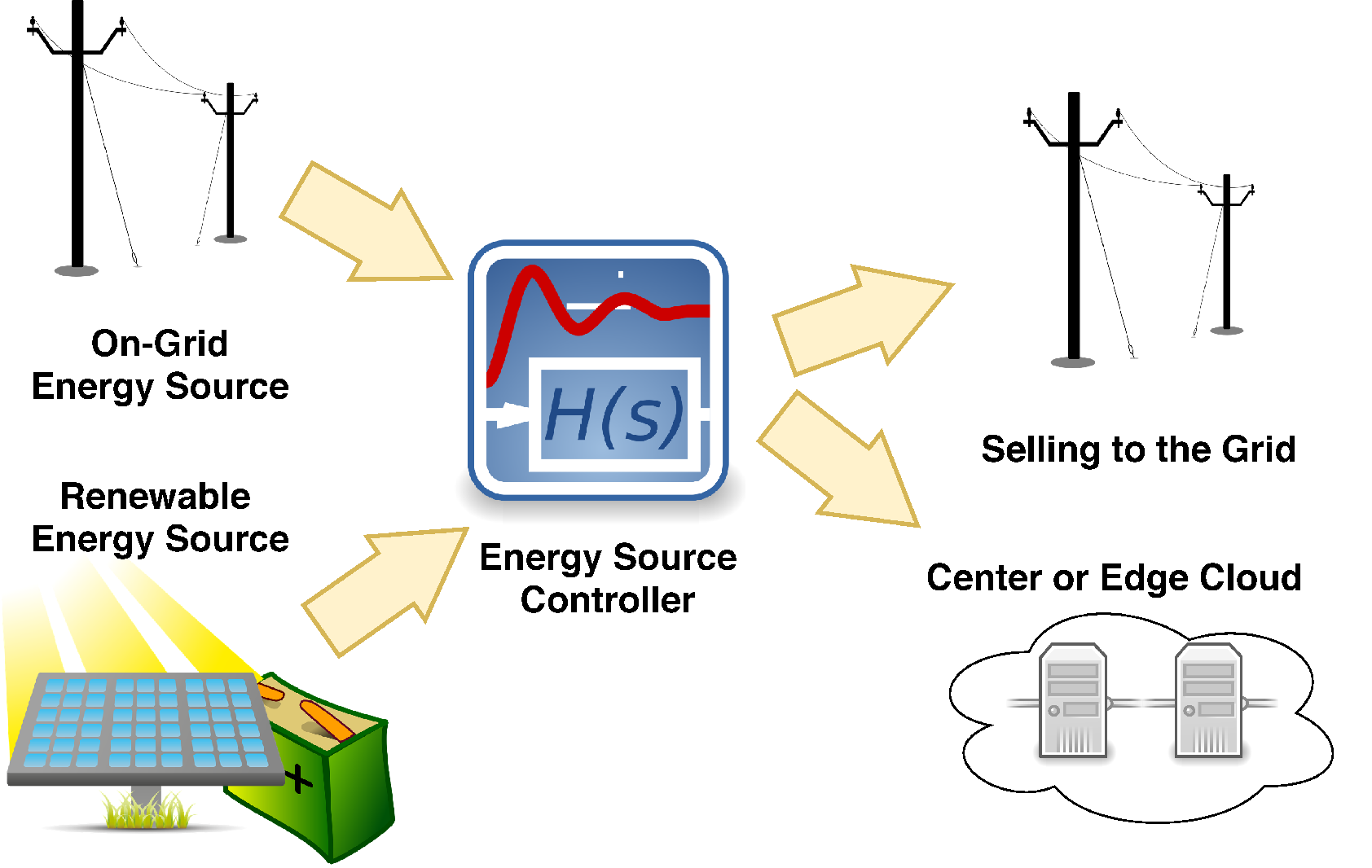}
\caption{\label{fig:rensysec} Energy Model in Green Hybrid CRAN.}
\end{figure}   
\subsection{Energy Model}
\par Figure~\ref{fig:rensysec} shows the electrical grid energy supported renewable energy system for a cloud station. In this model, a station has two different energy sources. In one side, the solar panel harvests the renewable energy from sunlight and the battery stores this energy. While that green energy reduces the brown energy consumption, the other energy source (on-grid) provides the energy requirement in the case of the lack of the insufficient green energy. Besides, using a pure green system is not economical and increases the TCO of an MNO \cite{Han2016}. According to Hassan et al. and Valerdi et al., harvested green energy may be used instantly by the system or may be stored in a battery for planned use \cite{Hassan2013a,valerdi2010intelligent}. Saving the green energy in a battery does prevent not only the wasting of the excess green energy but also advances the more economical use of green energy. Although this system is not common in a RAN nowadays, Bloomberg Finance group declared in one of their latest reports that using battery storage alongside with solar panels will become an ordinary method for a rooftop system in 2020 \cite{Parkinson2014}. Besides, a recent report about the solar panel technology in Germany also confirmed this new system. They imply that the prices of solar panels drop 19\% each year and they mention that the prices continue to drop year by year \cite{Wirth2016}. With this in mind, we directly focus on a cloud station which has their solar panels and batteries. 
\par The energy source controller, in the middle of Figure~\ref{fig:rensysec}, determines the energy source that is used to supply the cloud station. This component is critical for reducing the operational expenditure, which is explained in the next section. The energy consumption in a station has two elements; the first one is the static energy consumption which does not change by the amount of processing activity in this station such as cooling the system. The other one is the dynamic energy consumption which depends on the number of active digital units (DUs)\footnote{A digital unit (DU) is a processing unit in a cloud which process the functions of a classic BBU.} in this station. Equation~\ref{eq:energyConsInCS} and Equation~\ref{eq:energyConsInRS} calculate the total energy consumption in a CC and a EC, respectively. Lastly, in addition to the using renewable energy in a station, an MNO may sell the surplus renewable energy to the grid. 
\begin{equation}
\label{eq:energyConsInCS}
\Psi_{CCt} = \left(PS_{CC} + \sum\limits_{d\in\mathcal{D}_{CC}}a_{dt}PD_{CC}\right)
\end{equation}
\begin{equation}
\label{eq:energyConsInRS}
\Psi_{rt} = \left[PS_{EC} + \sum\limits_{d\in\mathcal{D}_{r}}a_{dt}PD_{EC} \right]
\end{equation}

\begin{table}
\centering
\caption{\label{tab:Notations} Sets \& Variables}
\begin{tabular}{|c|p{6.5cm}| }
\hline
Sets &  Explanation \\ \hline
$t\in\mathcal{T}$ & set of time intervals \\
$i\in\mathcal{I}$ & set of user equipments \\
$c\in\mathcal{C}$ & set of RRHs \\
$d\in\mathcal{D}$ & set of DUs \newline ($d=CC$ is the set of DUs in CC)\\
$r\in\mathcal{R}$ & set of ECs \\
$f\in\mathcal{F}$ & set of URFs \\ \hline
Variables&  Explanation \\ \hline
$m_{idft}$ & whether URF $f$ of UE i is hosted in DU $d$ \newline in time interval $t$ or not\\
$a_{dt}$ & whether DU $d$ is active in time interval $t$ or not \\
$s_{rt}$ &  green energy consumption \newline in EC $r$ in time interval $t$ \\
$p_{rt}$ &  sold energy \newline in EC $r$ in time interval $t$ \\
$b_{rt}$ &  green energy in the battery \newline of EC $r$ in time interval $t$ \\  \hline
\end{tabular}
\end{table} 
\begin{table}
\centering
\caption{\label{tab:constants}Input Variables}
\begin{tabular}{|c|p{6cm}| }
\hline
Power Cons. &  Explanation \\ \hline
$PD_{CC}$ & power consumption of a DU in CC\\
$PD_{EC}$ & power consumption of a DU in EC\\
$PS_{CC}$ & static power consumption in CC\\
$PS_{EC}$ &  static power consumption in a EC\\ \hline
Others &  Explanation \\ \hline
$\rho_{it}$ & traffic load ratio of user $i$\\
$\mu_{it}$ & delay threshold of user $i$\\
$L_{y}$ & DU function cap. ($y$ is $CC$ or $EC$)\\
$B_{r}$ & battery maximum storage capacity\\
$S_{r}$ & solar panel size\\
$G_{rt}$ & generated green energy in time interval $t$\\
$E_{t}$ & energy price in time interval $t$\\
$P$ & sold energy penalty ratio\\
\hline
\end{tabular}
\end{table} 

\section{Green-Aware Function Split Optimization Problem}

\par The following optimization problem minimizes the operational expenditure (OpEx) of an MNO which operates the Green Hybrid CRAN.  Table~\ref{tab:Notations} and Table~\ref{tab:constants} summarize the notations in this section.
\par Minimize:
\begin{equation}
\begin{split}
\label{eq:obj1}
\sum\limits_{t\in\mathcal{T}} & \biggl [  \Psi_{CCt}-s_{CCt}-P*p_{CCt}  \\
 & +  \sum\limits_{r\in\mathcal{R}}(\Psi_{rt}-s_{rt}-P*p_{rt}) \biggr] * E_{t}
\end{split}
\end{equation}
s.t.:
\begin{equation}
\label{eq:DUCapInCS}
\sum\limits_{f\in\mathcal{F}}\sum\limits_{i\in\mathcal{I}} \rho_{it} m_{idft} < L_{CC} ,  \forall d \in \mathcal{D}_{CC}, \forall t \in \mathcal{T}
\end{equation}
\begin{equation}
\label{eq:DUCapInRS}
\sum\limits_{f\in\mathcal{F}}\sum\limits_{c\in\mathcal{C}_{r}} \sum\limits_{i\in\mathcal{I}_{c}} \rho_{it}  m_{idft} < L_{RS}, \forall d \in \mathcal{D}_{r} ,  \forall r \in \mathcal{R}, \forall t \in \mathcal{T}
\end{equation}
\begin{equation}
\label{eq:DUActiveInCS}
M * a_{dt} -\sum\limits_{f\in\mathcal{F}}\sum\limits_{i\in\mathcal{I}}m_{idft}\geq 0 ,  \forall d \in \mathcal{D}_{CC},\forall t \in \mathcal{T}
\end{equation}
\begin{equation}
\label{eq:DUActiveInRS}
M * a_{dt} - \sum\limits_{f\in\mathcal{F}}\sum\limits_{c\in\mathcal{C}_{r}} \sum\limits_{i\in\mathcal{I}_{c}}m_{idft}\geq 0 ,  \forall d \in \mathcal{D}_{r}, \forall r \in \mathcal{R}, \forall t \in \mathcal{T}
\end{equation}
\begin{equation}
\label{eq:FunctionAssign}
\sum\limits_{f\in\mathcal{F}}\sum\limits_{d\in\mathcal{D}_{CC}\cup\mathcal{D}_{r}}m_{idft} = |\mathcal{F}|, \forall i\in\mathcal{I}_{c}, c\in\mathcal{C}_{r},  \forall r \in \mathcal{R}, \forall t \in \mathcal{T}
\end{equation}
\begin{equation}
\label{eq:DelayConst}
\sum\limits_{f\in\mathcal{F}} \sum\limits_{d\in\mathcal{D}_{CC}} m_{idft} < \mu_{it} ,  \forall i \in \mathcal{I}, \forall t \in \mathcal{T}
\end{equation}
\begin{equation}
\label{eq:battEnergyInCS}
b_{CCt} = b_{CC(t-1)} - s_{CCt} - p_{CCt} + S_{CC}G_{CCt},  \forall t \in \mathcal{T}
\end{equation}
\begin{equation}
\label{eq:battEnergyInRS}
b_{rt} = b_{r(t-1)} - s_{rt} - p_{rt} + S_{r}G_{rt}, \forall r\in\mathcal{R},  \forall t \in \mathcal{T}
\end{equation}
\begin{equation}
\label{eq:battLimitInCS}
 b_{CCt} \leq B_{CC},  \forall t \in \mathcal{T}
\end{equation}
\begin{equation}
\label{eq:battLimitInRS}
 b_{rt} \leq B_{r}, \forall r\in\mathcal{R},  \forall t \in \mathcal{T}
\end{equation}
\begin{equation}
\label{eq:renEnMaxLimitInCS}
s_{CCt} \leq  \Psi_{CCt},\forall t \in \mathcal{T}
\end{equation}
\begin{equation}
\label{eq:renEnMaxLimitInRS}
s_{rt} \leq  \Psi_{rt} ,\forall r\in\mathcal{R}, \forall t \in \mathcal{T}
\end{equation}
\par The objective function points that we have to consider three significant actions to reduce the cost of an MNO. The first one is reducing the total energy consumptions in the CC and ECs. We can achieve that by lowering the number of active DUs in these stations (clouds). The second one is increasing the usage of renewable energy in these stations. This action principally depends on the size of solar panels and the batteries in these stations and planning the use of renewable energy in these batteries in an efficient way. Alternatively, MNO may sell this valuable energy to the grid network at a reduced price \cite{Ahmed2018}. The third one is that we have to consider the changing electricity prices in a day period to reduce the overall cost of the MNO\footnote{We have to notice that this is an online problem, the bandwidth between the ECs and CC is already leased before the splitting decisions. Therefore, minimizing the bandwidth is not considered in this problem.}. The maintenance cost of this green system, which is another component of the operational expenditure, is not included in  Equation~\ref{eq:obj1}. The reason is that this maintenance cost is a constant value in this system and does not change by any decision variable in this equation.
\par DUs have limited capacities to execute URFs. Inequality~(\ref{eq:DUCapInCS}-\ref{eq:DUCapInRS}) show this limitation in CC and ECs, respectively. According to Mharsi et al., the processing requirement of user functions depend on the traffic loads \cite{Mharsi2018}. Therefore, we also add the traffic load of each user in these inequalities. 
\par As we mentioned before, we try to minimize the number of active DUs in CC and ECs. However, we have to activate a DU, if it is used for processing a user function. Inequality~(\ref{eq:DUActiveInCS}-\ref{eq:DUActiveInRS}) provide this relation for CC and ECs, respectively. Inequality~\ref{eq:FunctionAssign} guarantees another critical constraint which provides that DUs process all demanded URFs in the network. End-to-End delay is essential for quality of service and Inequality~\ref{eq:DelayConst} bounds it for each user demand in each time slot\footnote{The computing costs are assumed the same for each URF. Therefore, deciding the number of URFs in one cloud side provide us a certain splitting point in the chain of the URF.}.
\par The amount of the green energy we can use in a station depends on the remaining energy in the battery from the previous time slot, consumed green energy,  the sold energy to the grid and generated renewable energy. Inequalities~(\ref{eq:battEnergyInCS}-\ref{eq:battEnergyInRS}) provide this relation by calculating the remaining energy in the batteries of CC and ECs, respectively. Another important restriction of using green energy in the network is the size of batteries in CC and ECs which are shown by Inequalities~(\ref{eq:battLimitInCS},\ref{eq:battLimitInRS}). Lastly, it is clear that the consumed green energy should not be higher than the needed energy consumption in a station. Inequalities~(\ref{eq:renEnMaxLimitInCS},\ref{eq:renEnMaxLimitInRS}) provide this limitation for CC and ECs, respectively.
\par This problem involves the bin packing problem \cite{Alabbasi2018}; thus it is an NP-Hard problem, and we use a MILP Solver for up to certain size problems and provide a novel heuristic approach to find a solution for larger RANs. This heuristic is explained in the next section.
  
\section{A Heuristic Approach for Green Energy-Aware Function Split Optimization Problem}
\par
\begin{figure}
\centering
\includegraphics[width=0.48\textwidth]{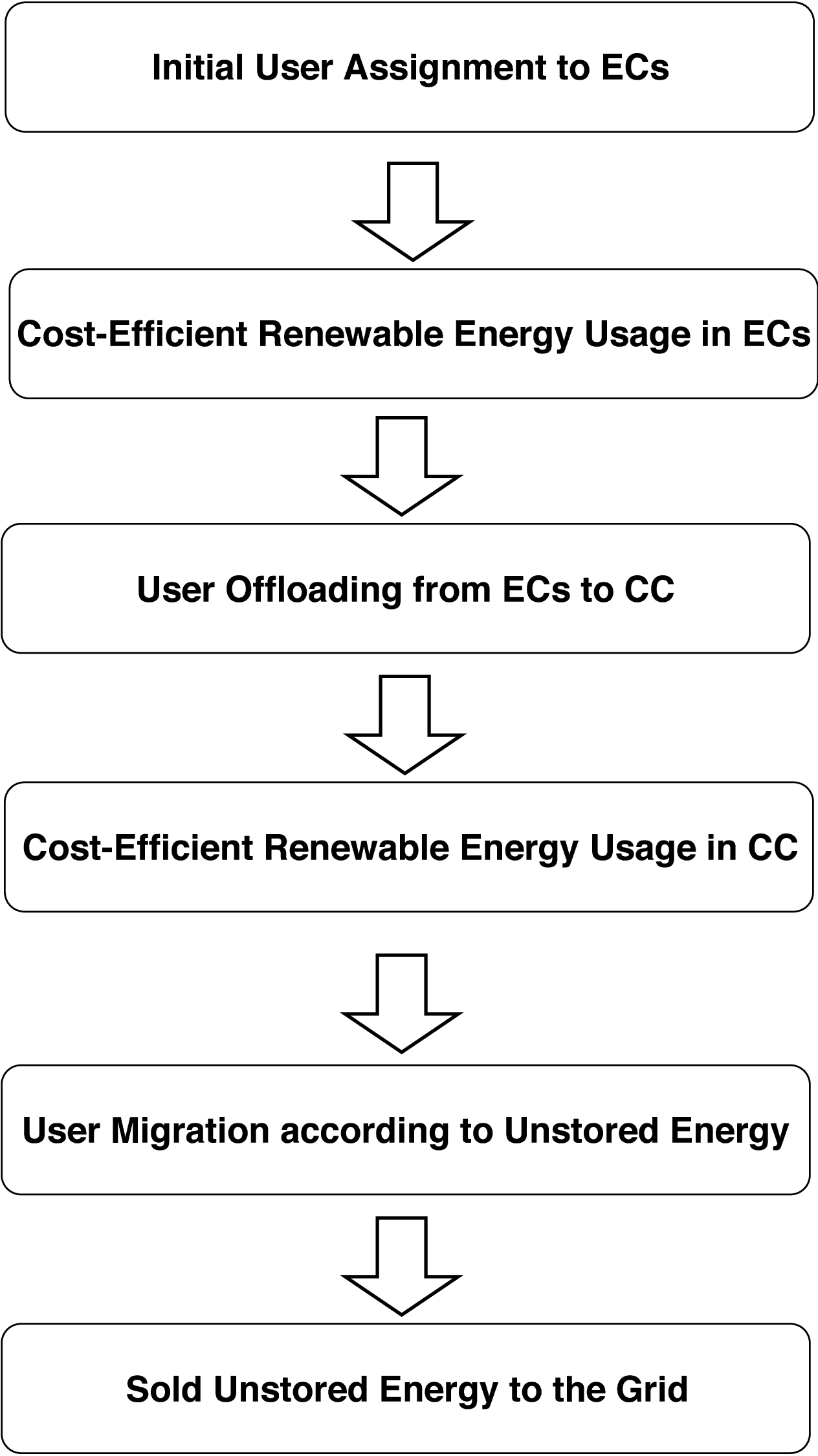}
\caption{\label{fig:heuristicmain} Heuristic Steps.}
\end{figure}

 The main steps of the heuristic are shown in Figure~\ref{fig:heuristicmain}. The input variables for the heuristic summarize in Table~\ref{tab:constants}. In the end of the heuristic, we find the objective function value (Equation \ref{eq:obj1}) and the decision variables (Table~\ref{tab:Notations}).
\begin{algorithm}
\caption{Initial User Assignment to ECs}
\label{alg:initialUserAssignment}
\begin{algorithmic}[1]
\STATE Given: $\rho, r$
\STATE $d = 0, \hat{LUD}=\hat{0}$
\FORALL{${i} \in I_{r}$}
\FORALL{${f} \in F$}
\IF {$\hat{LUD} + \rho_{i} \leq L_{EC}$}
\STATE $d = d + 1$
\ENDIF
\STATE $\hat{LUD}_{d} = \hat{LUD}_{d} + \rho_{i}$
\STATE $m_{idf} = 1$
\ENDFOR
\ENDFOR
\FORALL{${d} \in D_{r}$}
\FORALL{${i} \in I_{r}$}
\FORALL{${f} \in F$}
\IF{$m_{idf} = 1$}
\STATE $a_{d} = 1$
\ENDIF
\ENDFOR
\ENDFOR
\ENDFOR
\end{algorithmic}
\end{algorithm}
\par The heuristic starts with an initial assignment of the user functions to the ECs (Algorithm~\ref{alg:initialUserAssignment}). This algorithm runs for each edge cloud separately; thus we omit the cloud indices for the presentation simplicity. The notation $\hat{LUD}$ represents loads of the DUs. At the beginning of the algorithm, DUs do not serve any user function; thus loads of DUs are initialized with zero. Then, we assign the each user function to a DU in an ascending order ($m_{idf} = 1$). If a DU load gets full, we increase the index ($d$) and assign this user function to the next DU. After we finish the assignment of user functions to the DUs, we check the activity of the DUs in this cloud. If a DU serves any user function, we switch on this DU $a_{d}=1$.
\par In the second step, we make the decisions of using renewable energy for each ECs (Algorithm~\ref{alg:renewableEnergyUsage}). Notice once again that we omit the cloud indices for presentation simplicity.  The notation  $\hat{TEC}$ represents the total energy consumption, which is the summation of the grid and renewable energy consumptions in each time slot. The notation $\hat{RES}$ represents the amount of the reserved energy for each time slot. This vector variable is the fundamental concept of this algorithm which provides us to use the renewable energy in the most profitable time slot without violating the physical constraints of the solar panel and the battery in the renewable energy system of an EC. The algorithm starts with sorting the time slots by considering the electricity prices in these time slots. Then, by starting with the highest priced time slot, we check whether the $\hat{RES}$ is assigned before in this loop or not. If we have not assigned to the $\hat{RES}$ previously, available energy $availableEn$ equals to the generated renewable energy in this time slot. A previous assignment to the $\hat{RES}$ means that for a later time slot of a day, we need to reserve the renewable energy of this time slot; thus we restrict the usage of renewable energy in this situation. In the next step, we sum the available energy and the remaining energy in the battery to find the permitted renewable energy ($nextB$) that we can use in this time slot. Besides, between line 16-20, we restrict the usage of renewable energy by the total energy consumption for avoiding unnecessary renewable energy consumption. After deciding the amount of renewable energy use in this time slot ($s_{t}$), we update the remaining energy in the batteries for the later time slots in the lines between 21-23. Lastly, we update the reserved energy of the previous time slots of the day in the lines between 24-36. The notation $demandedEn$ represent the demanded renewable energy from the previous time slots. We subtract the generated energy ($G_{t}$) on each time slot if we have not used it previously. When the $demandedEn$ become non-positive, we stop the loop. In conclusion, with this algorithm, we can freely choose any time slot to use the renewable energy. The remaining energy $\hat{b}$ prevents the use of non-generated renewable energy in the following time slots and, $\hat{RES}$ precludes the use of reserved energy in the previous time slots. 
\begin{algorithm}
\caption{Cost-Efficient Renewable Energy Usage in a Cloud}
\label{alg:renewableEnergyUsage}
\begin{algorithmic}[1]
\STATE Given: $\hat{TEC}, \hat{b}, PrevDayEn$
\STATE $d = 0$, $\hat{LUD}=\hat{0}$
\STATE $\hat{RES} = MAGIC\_NUMBER$
\STATE $T^s= sort(\arg\max\limits_{t\in\mathcal{T}}(E_{t}))$
\FORALL{${t} \in T^{s}$}
\IF {$\hat{RES}_{t} = MAGIC\_NUMBER$}
\STATE $availableEn=G_{t}$
\ELSE
\STATE $availableEn = - \max(\hat{RES}_{t}, 0)$
\ENDIF
\IF{t=0}
\STATE $nextB=PrevDayEn+availableEn$
\ELSE
\STATE $nextB=b_{t-1}+availableEn$
\ENDIF
\IF{$nextB\geq \hat{TEC}_{t}$}
\STATE {$s_{t} =  \hat{TEC}_{t}$}
\ELSE 
\STATE {$s_{t} =  nextB$}
\ENDIF
\FOR{$t^{l}:=t$ \TO $\mid T \mid$}
\STATE {$Update(b_{t^{l}})$}
\ENDFOR
\IF{$t \neq 0$}
\STATE $demandedEn = max(s_{t} - G_{t}, 0)$
\FOR{$t^{p}:=t-1$ \TO $0$}
\IF {$\hat{RES}_{t^{p}} = MAGIC\_NUMBER$}
\STATE $\hat{RES}_{t^{p}} = demandedEn - G_{t^{p}} $
\STATE $demandedEn = demandedEn - G_{t^{p}} $
\ELSE
\STATE $\hat{RES}_{t^{p}} = \hat{RES}_{t^{p}} + demandedEn $
\ENDIF
\IF {$\hat{RES}_{t^{p}} \leq 0$}
\STATE BREAK
\ENDIF
\ENDFOR
\ENDIF
\ENDFOR
\end{algorithmic}
\end{algorithm}
\par In the third step, we offload the users from ECs to CC to reduce the grid energy consumption in ECs and to get benefits of the generated renewable energy in CC. Algorithm~\ref{alg:userOffloadingDecision} shows how to choose the appropriate user function to offload to the CC. We run this algorithm for each EC and the time slot separately. Thus, we omit the cloud and time slot notations in the algorithm for clarified presentation. The idea of this algorithm relies on reducing the number of active DUs in EC. In the first step of the heuristic 
(Algorithm~\ref{alg:initialUserAssignment}), we assign the user functions to the DUs of EC in an ascending order. Therefore, the highest indexed DUs tend to have lower traffic load ($\hat{LUD}$). Thus, in this algorithm ( Algorithm~\ref{alg:userOffloadingDecision}), we choose the user functions from the DUs in reverse order. In the loop section, we start to check whether the DU is active or not and if the EC consumes grid energy. If the EC consumes only green energy, we do not have to continue to migrate the user functions to the CC. Then for all user functions, we check that if this user function is assigned to the related DU. Lastly, we check the delay constraint. If we violate the delay constraint, we should not operate this migration. If the user function passes from all checks, we run the user offloading operation algorithm (Algorithm~\ref{alg:userOffloading}). First, we disconnect the user function from the DU in the EC. Then, we check if any user function is assigned to this DU. If this DU does not serve any user function, we switch it off to preserve the energy consumption. In the third step, we start to check each DU in CC in ascending order. If the DU capacity is enough to serve this new user function, we provide the assignment operations. Otherwise, we continue the checking operation by the next DU in CC.  After finishing all migrations in an EC, we check whether we reduce the number of active DUs in Line 22. If we fail to reduce it, we reverse all migrations to prevent impractical traffic loads in CC.
\begin{algorithm}
\caption{User Offloading Decision from ECs to CC}
\label{alg:userOffloadingDecision}
\begin{algorithmic}[1]
\STATE Given: $\hat{m},\hat{a}, \hat{LUD}$
\STATE $\hat{m}^P = \hat{m}$ 
\STATE $\hat{a}^P = \hat{a}$ 
\STATE $\hat{LUD}^P = \hat{LUD}$ 
\STATE $ NDU^P = \sum(m_{if}) $  
\FOR{$d:=\mid D_{r} \mid $ \TO $0$}
\IF {$a_{d} \neq 0$}
\IF{$GridConsumptionInEC$}
\FORALL{${i} \in I_{r}$}
\FORALL{${f} \in F$}
\IF{$m_{idf} = 1$}
\IF{$DelayCheckOk$}
\STATE $UserOffloading(i, d, f)$
\ELSE
\STATE BREAK
\ENDIF
\ENDIF
\ENDFOR
\ENDFOR
\ENDIF
\ENDIF
\IF{$ NDU^P \leq \sum(m_{if}) $ } 
\STATE $\hat{m} = \hat{m}^P$ 
\STATE $\hat{a} = \hat{a}^P$ 
\STATE $\hat{LUD} = \hat{LUD}^P$ 
\ENDIF
\ENDFOR
\end{algorithmic}
\end{algorithm}
\begin{algorithm}
\caption{User Offloading Operation from ECs to CC}
\label{alg:userOffloading}
\begin{algorithmic}[1]
\STATE Given: $m, \hat{LUD}, i, d, f$
\STATE $m_{idf} = 0$
\IF{ $\sum\limits_{i\in\mathcal{I}} \sum\limits_{f\in\mathcal{F}} m_{idf} = 0 $}
\STATE $a_{d} = 0$
\ENDIF
\FOR{$d_{CC}:=0$ \TO $\mid D_{CC} \mid $}
\IF {$\hat{LUD}_{d_{CC}} + \rho_{i} \leq L_{CC}$}
\STATE $\hat{LUD}_{d_{CC}} = \hat{LUD}_{d_{CC}} + \rho_{i}$
\STATE $m_{id_{CC}f} = 1$
\STATE $a_{d_{CC}} = 1$
\STATE BREAK
\ENDIF
\ENDFOR
\end{algorithmic}
\end{algorithm}
\par In the fourth step of the heuristic (Figure~\ref{fig:heuristicmain}), we provide a cost-efficient renewable energy usage in CC. This algorithm is the same algorithm we use for the ECs (Algorithm~\ref{alg:renewableEnergyUsage}). In the fifth step, we make migration of the user functions in two-way directions (ECs to CC and CC to ECs) according to unstored renewable energy in the batteries of the clouds. If only one side has unstored energy, we migrate the user functions from the other one to this side. 
\begin{algorithm}
\caption{User Migration Decision between the ECs and CC}
\label{alg:useMigrationDecision}
\begin{algorithmic}[1]
\STATE Given: $\hat{UnstoredEn}$
\FORALL{$t \in T$}
\IF{$\hat{UnstoredEn}_{CCt} > 0$}
\FORALL{$r\in R$}
\IF {$\hat{UnstoredEn}_{rt} = 0$}
\FORALL{$i\in I$ and $f\in F$}
\STATE $UserOffloading(i, d, f)$
\IF{$\hat{UnstoredEn}_{CCt} = 0$}
\STATE BREAK
\ENDIF
\ENDFOR
\ENDIF
\ENDFOR
\ELSE
\FORALL{$r\in R$}
\IF {$\hat{UnstoredEn}_ {rt} > 0$}
\FORALL{$i\in I$ and $f\in F$}
\STATE $UserOffloading(i, d, f)$
\IF{$\hat{UnstoredEn}_{rt} = 0$}
\STATE BREAK
\ENDIF
\ENDFOR
\ENDIF
\ENDFOR
\ENDIF
\ENDFOR
\end{algorithmic}
\end{algorithm}
\par Finally, in the last step of the heuristic (Figure~\ref{fig:heuristicmain}), we calculate the sold energy to the grid. This step has a trivial algorithm which directly sell the unstored renewable energy in the batteries of each cloud in each time slot. In the next section, we compare this heuristic with two MILP models that we also create to find the solutions for our optimization problem.
\section{Case Study and Results}
\begin{figure}
\centering
\includegraphics[width=0.48\textwidth]{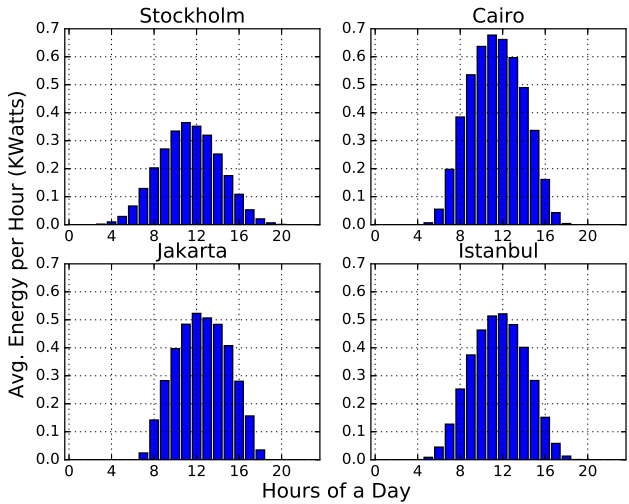}
\caption{\label{fig:harvested_hourly} Distribution of Harvested Solar Radiation in a Day Period.}
\end{figure}  
\begin{figure}
\centering
\includegraphics[width=0.48\textwidth]{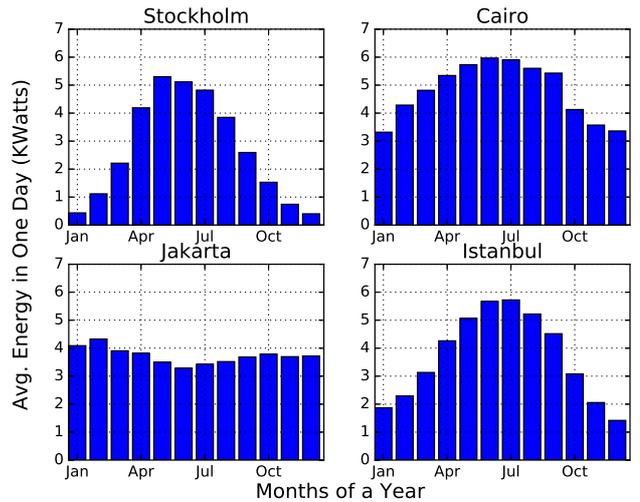}
\caption{\label{fig:harvested_daily} Distribution of Harvested Solar Radiation in a Year Period.}
\end{figure}
\par We test the system for different solar radiation distributions. We use the empirical data from the pvWatts application to calculate the amount of the generated energy of a solar panel for each time interval \cite{NationalRenewableEnergyLaboratory}. Their 30 years of historical weather data provide us to calculate the detailed solar energy generation rate data of a panel in a cloud station ($G_{rt}$) for four different cities. They provide each hour of the day data; thus we can simulate the change of the solar energy in several time scales. Figure~\ref{fig:harvested_hourly} and Figure~\ref{fig:harvested_daily} show the change of generated energy of a $4kW$ size of a solar panel for different cities. The delay constraints for each user ($\mu_{it}$) are chosen randomly, uniformly distributed on [0, $\mid\mathcal{F}\mid$]. Table~\ref{tab:Parameters} shows the other test parameters.
\par In our experimental setup, we study with 20 ECs and 1 CC. Each EC serves 8 RRHs, and each RRH serves 5, 10 or 15 demand points, which are called "Low", "Medium" and "High" traffic in the following results. Therefore, we can test the proposed heuristic and MILP solutions with different size of RANs. These demand points may represent the user groups that demand the data traffic from an RRH.
\begin{table}
\centering
\caption{\label{tab:Parameters} Experiment Parameters}
\begin{tabular}{|c|c| }
\hline
Explanation  & Value \\ \hline
$PD_{CC}$  &1500 W/h \\
$PD_{EC}$  &500 W/h \\
$PS_{CC}$  &750 W/h \\
$PS_{EC}$  &250 W/h \\
\hline
$L_{CC}$ & 50 URFs\\
$L_{EC}$ & 15 URFs\\
\hline
$B_{CC}$ & 20 KW/h\\
$B_{EC}$ & 5 KW/h\\
$S_{CC}$ & 20 KW/h\\
$S_{EC}$ & 5 KW/h\\
\hline
$E_{t}$ & [0.29, 0.46, 0.70]\footnotemark[2]\\
$P$&0.5 \\
\hline
\end{tabular}
\end{table}
\footnotetext[2]{The pricing data changes frequently and there is a significant difference between each city. The pricing data in our experiments comes from EPDK \cite{EPDK2018}. We use different prices according to the time of the day.}
\begin{figure}
\centering
\includegraphics[width=0.48\textwidth]{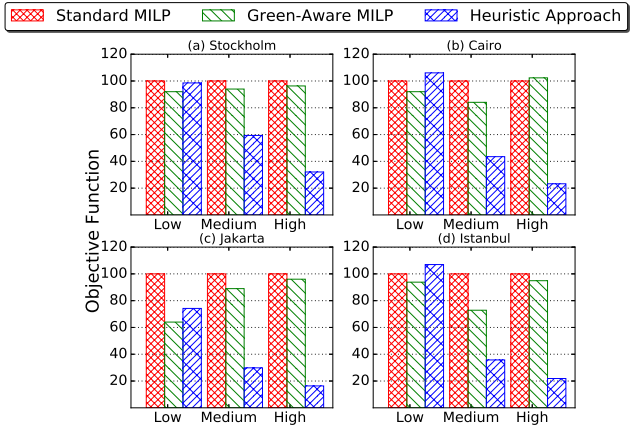}
\caption{\label{fig:bc} Operational Expenditure Comparison of Heuristic, Green Energy-Aware MILP and Standard Method}
\end{figure}  
\par Besides the heuristic we explain in the previous section, we use a Mixed Integer Linear Programming (MILP) Solver, Gurobi \cite{GurobiOptimization2018} to solve the green energy-aware function split optimization problem. Computation experiments were run on a Nvidia DGX-1 Station \cite{Nvidia2018} with a Dual 20-Core Intel Xeon E5-2698 v4 2.2 GHz. The termination time is chosen as 4 hours. We compare the MILP solution and the heuristic with a standard method in which energy source controller use the renewable energy source whenever it is available and does not determine the splitting decisions by considering renewable energy.
\par First, we have to explain the performance of the solver for that size of a problem. Gurobi finds the optimum solution only in low traffic rate and for only standard MILP model. Green energy-aware MILP finds the solution around 5\% gap from the lower bound in low traffic rate. In medium and higher traffic rates gap from the lower bound were higher than the 30\% for each model. The reason is that in the higher traffic rates and the Green MILP model, the solution space expands exponentially and 4 hours solution time limit is not enough to reach the optimum solution. Besides, we deal with an online problem, and we have to find a solution in an hour for a real scenario. Hence, we propose a heuristic to find a faster solution.
\begin{figure}
\centering
\includegraphics[width=0.48\textwidth]{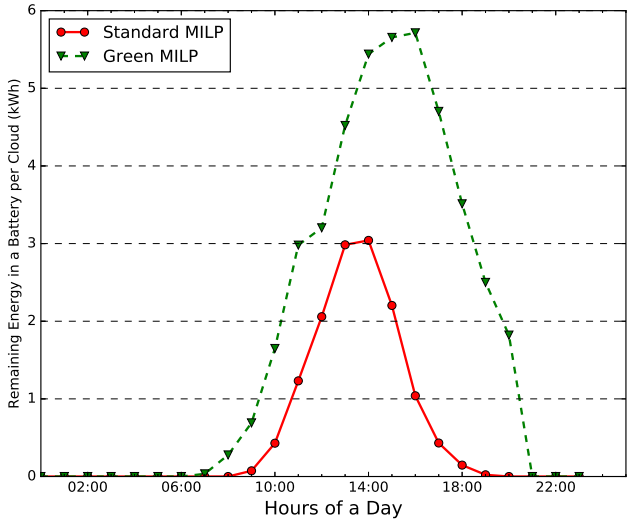}
\caption{\label{fig:remEn}Remaining Energy in the Batteries of Cloud Stations (Jakarta Medium Traffic)}
\end{figure}
\par Figure~\ref{fig:bc} shows the comparison of each method. If we compare two MILP solutions, it is obvious that green energy-aware splitting decisions provide better results for most of the traffic rates and solar radiations we studied. There are two reasons for that outcome and Figure~\ref{fig:remEn} demonstrates them. This figure shows the variation of the average remaining energy in the batteries of the cloud stations in a day period for Jakarta and medium traffic rate configuration. The first advantage of green-aware MILP is that it promotes using the remaining energy in more profitable hours. Second, it provides a better load balance between the CC and ECs, deciding the splitting decisions by promoting the station which has more renewable energy. Thus, it prevents the MNO to sell its renewable energy in a low profitable value.
\begin{figure}
\centering
\includegraphics[width=0.48\textwidth]{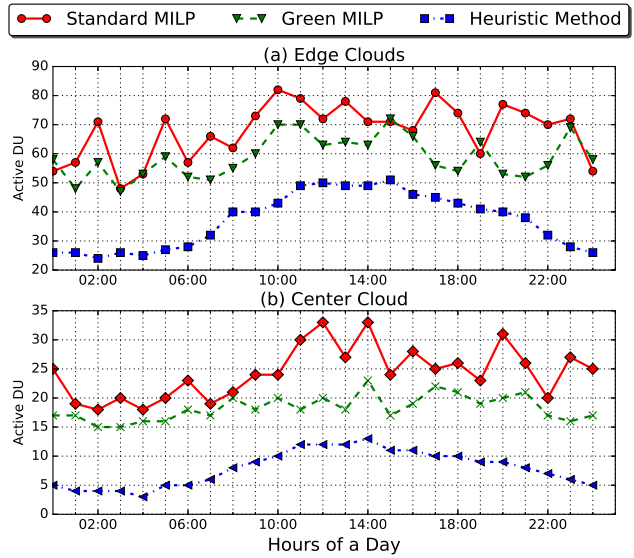}
\caption{\label{fig:du}Number of Active DUs in Each Cloud in a day period (Istanbul Medium Traffic).}
\end{figure}
\par If we now compare our proposed heuristic and the MILP solutions, Figure~\ref{fig:bc} clearly states that our heuristic approach provides outstanding results for medium and high traffic rates. Figure~\ref{fig:du} points out that outcome. This figure shows the number of active DUs for medium traffic rate in Istanbul. We may recognize that the number of active DUs in the heuristic solution is lower than both MILP solutions for CC and ECs. Thus, with this solution, the ECs and the CC consume less grid energy. 
\begin{figure}
\centering
\includegraphics[width=0.48\textwidth]{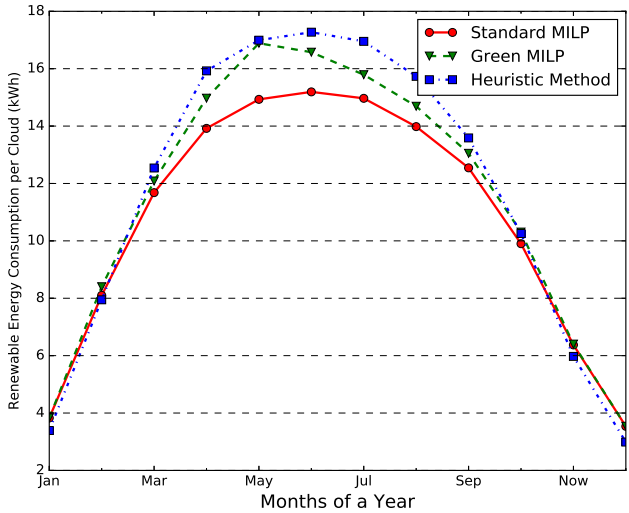}
\caption{\label{fig:rencos}Distribution of Renewable Energy Consumption in a year period (Stockholm Low Traffic).}
\end{figure}

\par We also analyze the results in a year period. Figure~\ref{fig:rencos} shows the consumption of renewable energy in a year period. Even, both solutions have the same size of solar panels and the batteries; the heuristic approach provides the higher level of renewable energy consumption especially in summer seasons in Stockholm. This result validates that the heuristic approach is better for promoting renewable energy consumption even in lower traffic rates. 
 
\par
\begin{figure}
\centering
\includegraphics[width=0.48\textwidth]{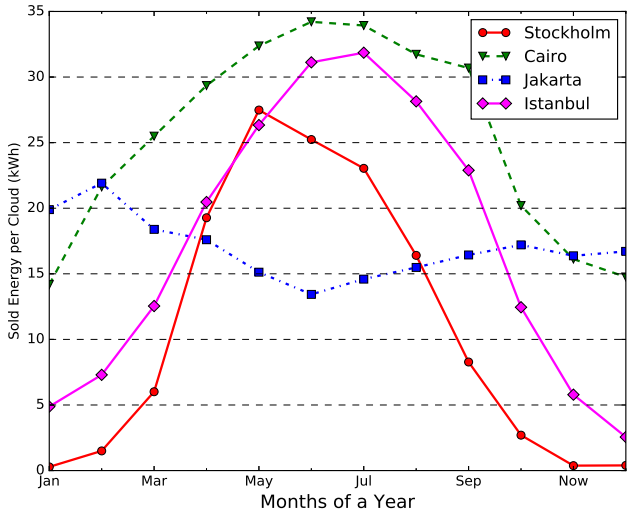}
\caption{\label{fig:sold}Distribution of the Sold Energy in a year period.}
\end{figure}

 Lastly, Figure~\ref{fig:sold} demonstrates the distribution of sold energy in a year period for different cities. Stockholm and Istanbul, which have four seasons, has a significant variation between the different seasons. The sold energy is higher in the summer season, and this energy can be used for cooling systems of cloud centers instead of selling at a low price for these cities.
 
\section{Conclusion}
\par A pure CRAN needs an enormous bandwidth capacity between an RRH and a BBU. Thus, the splitting of the BBU functions are proposed in recent studies. This paper develops these studies by adding renewable energy sources in CC and ECs. We explain the network architecture, the traffic, and energy models of this novel system. We formulate an operational expenditure minimization problem which decides the splitting options by considering to increase the renewable energy consumption, reducing the number of active DUs, and balancing the URFs between the stations and between the time slots. The results show that our proposed model which considers the renewable energy amount in the batteries reduces more operational expenditure and provides more profit to the MNOs. 
\par  In addition, this problem is an online problem, and an MNO should solve it on a daily basis. Thus, we proposed a fast heuristic, and the results showed that it provides an exceptional solution for large RANs. As future work, we are planning to improve the performance of our heuristic for small networks. Also, we will investigate the green-aware RRHs and their implementation to our current problem.  

\def\bibsection{\section*{References}}
\bibliography{der_hcran}

\end{document}